\documentclass[aps,pre,amsmath,amssymb,reprint,groupedaddress,showpacs]{revtex4-1}

\usepackage{graphicx}
\usepackage{dcolumn}
\usepackage{bm}

\begin{document}


\title{Electromagnetic space-time crystals. I. Fundamental solution of the Dirac equation}


\author{G. N. Borzdov}
\email[]{BorzdovG@bsu.by}
\affiliation{Department of Theoretical Physics and Astrophysics, Belarus State University,
Nezavisimosti avenue 4, 220030 Minsk, Belarus}



\begin{abstract}
The fundamental solution of the Dirac equation for an electron in an electromagnetic field with harmonic dependence on space-time coordinates is obtained. The field is composed of three standing plane harmonic waves with mutually orthogonal phase planes and the same frequency. Each standing wave consists of two eigenwaves with different complex amplitudes and opposite directions of propagation. The fundamental solution is obtained in the form of the projection operator defining the subspace of solutions to the Dirac equation.
\end{abstract}


\maketitle

\section{Introduction}
Considerable recent attension has been focussed on the possibility of time and space-time crystals~\cite{qtime,cltime,ions,gaps}, analogous to ordinary crystals in space. The papers~\cite{qtime,cltime} provide the affirmative answer to the question, whether time-translation symmetry might be spontaneously broken in a closed quantum-mechanical system~\cite{qtime} and a time-independent, conservative classical system~\cite{cltime}. A space-time crystal of trapped ions and a method to realize it experimentally by confining ions in a ring-shaped trapping potential with a static magnetic field is proposed in~\cite{ions}. Standing electromagnetic waves comprize another type of space-time crystals. It was shown~\cite{gaps} that one can treat the space-time lattice, created by a standing plane electromagnetic wave, by analogy with the crystals of nonrelativistic solid state physics. In particular, the wave functions, calculated within this framework by using the first-order perturbation theory for the Schr\"{o}dinger-Stuekelberg equation, are Bloch waves with energy gaps~\cite{gaps}.

Standing electromagnetic waves constitute an interesting family of localized fields which may have important practical applications. In particular, optical standing waves can be used to focus atoms and ions onto a surface in a controlled manner, nondiffracting Bessel beams can be used as optical tweezers which are noninvasive tools generating forces powerful enough to manipulate microscopic particles. Superpositions of homogeneous plane waves propagating in opposite directions, the so-called Whittaker expansions, play a very important role in analyzing and designing localized solutions to various homogeneous partial differential equations \cite{Don92,*Don93,*Sha95}.

In \cite{pre00,*pre01,pre02} we have proposed an approach to designing localized fields, that provides a broad spectrum of fields to construct electromagnetic fields with a high degree of two-dimensional and tree-dimensional spatial localization (2D and 3D localized fields) and promising practical applications. In particular, it can be used in designing fields to govern motions of charged and neutral particles. Some illustrations for relativistic electrons in such localized fields have been presented in~\cite{pre02}.

In this series of papers we treat the motion of the Dirac electron in an electromagnetic field with four-dimensional periodicity, i.e., with periodic dependence on all four space-time coordinates. In terms of the three-dimensional description, such electromagnetic space-time crystal (ESTC) can be treated as a time-harmonic 3D standing wave. In solid state physics, the motion of electrons in natural crystals is described by the Schr\"{o}dinger equation with a periodic electrostatic scalar potential. The description of the motion of electrons in ESTCs by the Dirac equation takes into account both the space-time periodicity of the vector potential and the intrinsic electron properties (charge, spin, and magnetic moment). In  this case, the Dirac equation reduces to an infinite system of matrix equations. To solve it, we generalize the operator methods developed in~\cite{cr1,*cr2,*cr3,*oc92,*jmp93,*jmp97,*Wiley} to the cases of infinite-dimensional spaces and finite-dimensional spaces with any number of space dimensions. The evolution, projection and pseudoinverse operators are of major importance in this approach. The evolution operator (the fundamental solution of a wave equation) describes the field dependence on the space-time coordinates for the whole family of partial solutions. The method of projection operators is very useful at problem solving in classical and quantum field theory~\cite{Fed58,Fed,Bogush}. It was developed by Fedorov~\cite{Fed58,Fed} to treat finite systems of linear homogeneous equations. In the frame of Fedorov's approach, it is necessary first to find projection operators which define subspaces of solutions for two subsystems (constituent parts) of the system to solve, and then to find its fundamental solution, i.e., the projection operator defining the intersection of these subspaces, by calculating the minimal polynomial for some Hermitian matrix of finite dimensions. In this paper, we present a different approach, based on the use of pseudoinverse operators, which is applicable to both finite and infinite systems of equations and has no need of minimal polynomials.

In this paper basic equations in matrix and operator forms are presented in Sec.~I. The projection operator of a system of homogeneous linear equations (the key concept of our approach) and the relations for its calculations by a recurrent algorithm are introduced in Sec.~II. For the infinite system under consideration, the appropriate recurrent algorithm and the fundamental solution are presented in Sec.~III. It is well known, e.g., see Ref.~\cite{Axi}, that 16 Dirac matrices form a basis in the space of $4\times 4$ matrices. In appendix, we introduce a specific numeration of these basis matrices, which makes it possible, in particular, to reconstruct the matrix from its number. It yields a convenient way both to represent derived matrix expressions in a concise form and accelerate numerical calculations. The subsequent two papers will present a fractal approach to numerical analysis of electromagnetic crystals and some results of this analysis, respectively.

\section{\label{baseq}Basic system of equations}
\subsection{Matrix form}
An electron in an electromagnetic field with the four-dimensional potential $\bm{A}=(\textbf{A},i \varphi)$ is described by the Dirac equation
\begin{equation}\label{Dirac}
   \left[\gamma_k\left(\frac{\partial}{\partial x_k} - i A_k\frac{e}{c \hbar}\right) + \kappa_{e}\right]\Psi=0,
\end{equation}
where $\kappa_{e}=m_{e} c/ \hbar$, $c$ is the speed of light in vacuum, $\hbar$ is the Planck constant, $e$ is the electron charge, $m_{e}$ is the electron rest mass,  $\gamma_k$ are the Dirac matrices, $\Psi$ is the bispinor, $x_1$, $x_2$ and $x_3$  are the Cartesian coordinates, $x_4=i c t$, and summation over repeated indices is carried out from 1 to 4. The exact solution of this equation for the Dirac electron in the field of a plane electromagnetic wave was obtained in a few forms, e.g., see~\cite{Fed,Tern}, and a general approach to find approximate solutions of Dirac type equations for particles of any spin in a weak electromagnetic field with an arbitrary dependence on space-time coordinates was suggested by Fedorov~\cite{Fed}. In this paper, we treat the field with $A_4\equiv i \varphi=0$ and
\begin{equation}\label{field}
  \textbf{A}^\prime \equiv \frac{e}{m_e c^2}\textbf{A}=\sum_{j=1}^6\left(\textbf{A}_j e^{i \bm{K}_j \cdot \bm{x}}+\textbf{A}_j^\ast e^{-i \bm{K}_j \cdot \bm{x}}\right),
\end{equation}
which is composed of six plane waves with unit wave normals $\pm\textbf{e}_\alpha$, where $\textbf{e}_\alpha$ are the orthonormal basis vectors, $\alpha =1, 2, 3$; $\bm{x}=(\textbf{r},ict)$, $\textbf{r}=x_1\textbf{e}_1+x_2\textbf{e}_2+x_3\textbf{e}_3$. All six waves
have the same frequency $\omega_0$ and
\begin{eqnarray}
\bm{K}_j&=&k_0 \bm{N}_j, \quad j=1,2,...,6, \quad k_0=\frac{\omega_0}{c}=\frac{2\pi}{\lambda_0},\nonumber
\\
\bm{N}_j&=&(\textbf{e}_j,i), \quad \bm{N}_{j+3}=(-\textbf{e}_j,i), \quad j=1,2,3. \label{KNj}
\end{eqnarray}
They may have any polarization, so that their complex amplitudes are specified by dimensionless real constants  $a_{jk}$ and $b_{jk}$ as follows
\begin{equation}
    \textbf{A}_j = \sum_{k=1}^3 \left(a_{jk} + i b_{jk}\right)\textbf{e}_k, \quad j = 1,2,...,6,\label{abjk}
\end{equation}
where $a_{jj} = b_{jj} = a_{j+3\,j} = b_{j+3\,j} = 0, j=1,2,3.$

We seek the solution of the Dirac equation in the form of a Fourier series
\begin{equation}\label{sol1}
    \Psi(\bm x)=\sum_{n \in \mathcal L} c(n) e^{i [\bm{K} + \bm{G}(n)]\cdot \bm{x}},
\end{equation}
where $\bm{K} = (\textbf{k},i \omega /c)$ is the four-dimensional wave vector, $\textbf{k} = k_1\textbf{e}_1+k_2\textbf{e}_2+k_3\textbf{e}_3$, $n=(n_1,n_2,n_3,n_4)$ is the multi-index specifying  $\textbf{n} = n_1\textbf{e}_1+n_2\textbf{e}_2+n_3\textbf{e}_3$ and $\bm{G}(n) = (k_0 \textbf{n},i k_0 n_4)$. Here, $c(n)$ are the Fourier amplitudes (bispinors), and $\mathcal L$ is the infinite set of all multi-indices $n$ with an even value of the sum $n_1+n_2+n_3+n_4$. Substitution of $\textbf{A}$ (\ref{field}) and $\Psi$ (\ref{sol1}) in Eq.~(\ref{Dirac}) results in the infinite system of matrix equations
\begin{equation}\label{meq}
\sum_{s \in S_{13}} V(n,s)c(n+s)=0, \quad {n\in \mathcal{L}},
\end{equation}
where $S_{13} = \{s_h(i),i=0,1,...,12\}$ is the set of 13 values of function $s_h=s_h(i)$, where $s_h(0)=(0,0,0,0)$. At $i=1,...,12$, this function specifies the shifts $s=(s_1,s_2,s_3,s_4)=s_h(i)$ of multi-indices $n$, defined by the Fourier spectrum of the field $\textbf{A}$ (\ref{field}), which satisfy the condition $|s_1|+|s_2|+|s_3|=|s_4|=1$. Because of this, they will be denoted the shifts of the first generation [$g_{4d}(s)=1$]. By the definition, $g_{4d}(s_1,s_2,s_3,s_4)=\max\{|s_1|+|s_2|+|s_3|,|s_4|\}$. Thus, each equation of the system relates 13 Fourier amplitudes (bispinors), in other words, each amplitude enters in 13 different matrix equations.

Due to the structure of the Dirac equation, the expansion of $4\times 4$ matrices in the basis formed by 16 Dirac matrices ($\Gamma_k, k=0,...,15$) yields a convenient way both to represent derived matrix expressions in a concise form and accelerate numerical calculations (see appendix). To this end, $4\times 4$ matrix $V$ is described by the set of its components in the Dirac basis [Dirac set of matrix $V$, briefly, D-set of $V$, or $D_s(V)$]. Let us introduce the dimensionless parameters
\begin{equation}\label{Qw}
    \bm Q = (\textbf{q},i q_4) = {\bm K}/\kappa_e, \quad  \Omega = \frac{\hbar \omega_0}{m_e c^2},
\end{equation}
\begin{equation}\label{qq4}
\textbf{q} = q_1\textbf{e}_1+q_2\textbf{e}_2+q_3\textbf{e}_3 = \frac{\hbar \textbf{ k}}{m_e c}, \quad   q_4 = \frac{\hbar \omega}{m_e c^2}.
\end{equation}
In this notation, the matrix coefficients $V[n,s_h(i)]$, in order of increasing $i=0,1,...,12$, have the following D-sets:
\begin{widetext}
\begin{eqnarray}
D_s\{V[n,(0,0,0,0)]\}&=&\{1,0,0,0,-w_4,0,0,0,0,0,0,0,0,i w_3,i w_1,i w_2\},\nonumber\\
D_s\{V[n,(0,0,-1,-1)]\}&=&\{0,0,0,0,0,0,0,0,0,0,0,0,0,0,-i a_{31}+b_{31},-i a_{32}+b_{32}\},\nonumber\\
D_s\{V[n,(0,-1,0,-1)]\}&=&\{0,0,0,0,0,0,0,0,0,0,0,0,0,-i a_{23}+b_{23},-i a_{21}+b_{21},0\},\nonumber\\
D_s\{V[n,(-1,0,0,-1)]\}&=&\{0,0,0,0,0,0,0,0,0,0,0,0,0,-i a_{13}+b_{13},0,-i a_{12}+b_{12}\},\nonumber\\
D_s\{V[n,(1,0,0,-1)]\}&=&\{0,0,0,0,0,0,0,0,0,0,0,0,0,-i a_{43}+b_{43},0,-i a_{42}+b_{42}\},\nonumber\\
D_s\{V[n,(0,1,0,-1)]\}&=&\{0,0,0,0,0,0,0,0,0,0,0,0,0,-i a_{53}+b_{53},-i a_{51}+b_{51},0\},\nonumber\\
D_s\{V[n,(0,0,1,-1)]\}&=&\{0,0,0,0,0,0,0,0,0,0,0,0,0,0,-i a_{61}+b_{61},-i a_{62}+b_{62}\},\nonumber\\
D_s\{V[n,(0,0,-1,1)]\}&=&\{0,0,0,0,0,0,0,0,0,0,0,0,0,0,-i a_{61}-b_{61},-i a_{62}-b_{62}\},\nonumber\\
D_s\{V[n,(0,-1,0,1)]\}&=&\{0,0,0,0,0,0,0,0,0,0,0,0,0,-i a_{53}-b_{53},-i a_{51}-b_{51},0\},\nonumber\\
D_s\{V[n,(-1,0,0,1)]\}&=&\{0,0,0,0,0,0,0,0,0,0,0,0,0,-i a_{43}-b_{43},0,-i a_{42}-b_{42}\},\nonumber\\
D_s\{V[n,(1,0,0,1)]\}&=&\{0,0,0,0,0,0,0,0,0,0,0,0,0,-i a_{13}-b_{13},0,-i a_{12}-b_{12}\},\nonumber\\
D_s\{V[n,(0,1,0,1)]\}&=&\{0,0,0,0,0,0,0,0,0,0,0,0,0,-i a_{23}-b_{23},-i a_{21}-b_{21},0\},\nonumber\\
D_s\{V[n,(0,0,1,1)]\}&=&\{0,0,0,0,0,0,0,0,0,0,0,0,0,0,-i a_{31}-b_{31},-i a_{32}-b_{32}\},\label{DsV}
\end{eqnarray}
\end{widetext}
where $n=(n_1,n_2,n_3,n_4)$, $w_j=q_j+n_j \Omega$.

\subsection{Operator form}
Let us treat the infinite set $C =\{c(n),n \in {\mathcal L}\}$ of the Fourier amplitudes $c(n)$ of the wave function $\Psi$ (\ref{sol1}) as an element of an infinite dimensional linear space $V_C$. Since, for any $n \in {\mathcal L}$,
\begin{equation}\label{cn}
    c(n) = \left(
             \begin{array}{c}
               c^1(n) \\
               c^2(n) \\
               c^3(n) \\
               c^4(n) \\
             \end{array}
           \right) \equiv \left(
           \begin{array}{c}
               c^1 \\
               c^2 \\
               c^3 \\
               c^4 \\
             \end{array}
           \right)_n
\end{equation}
is the bispinor, $C \in V_C$ will be denoted the multispinor. Let us define a basis $e_j(n)$ in $V_C$ and the dual basis $\theta^j(n) = e_j^{\dag}(n)$ in the space of one-forms $V_C^\ast$ ($n \in {\mathcal L}$):
\begin{eqnarray}\label{e14n}
  e_1(n)&=&\left(
             \begin{array}{c}
               1 \\
               0 \\
               0 \\
               0 \\
             \end{array}
           \right)_n, \quad e_2(n) = \left(
             \begin{array}{c}
               0 \\
               1 \\
               0 \\
               0 \\
             \end{array}
           \right)_n, \nonumber\\
  e_3(n)&=&\left(
             \begin{array}{c}
               0 \\
               0 \\
               1 \\
               0 \\
             \end{array}
           \right)_n, \quad e_4(n) = \left(
             \begin{array}{c}
               0 \\
               0 \\
               0 \\
               1 \\
             \end{array}
           \right)_n,
\end{eqnarray}
\begin{eqnarray}\label{t14n}
  \theta^1(n)&=&\left(
                          \begin{array}{cccc}
                            1 & 0 & 0 & 0 \\
                          \end{array}
                        \right)_n, \quad \theta^2(n) = \left(
                          \begin{array}{cccc}
                            0 & 1 & 0 & 0 \\
                          \end{array}
                        \right)_n,\nonumber  \\
  \theta^3(n)&=&\left(
                          \begin{array}{cccc}
                            0 & 0 & 1 & 0 \\
                          \end{array}
                        \right)_n, \quad \theta^4(n) = \left(
                          \begin{array}{cccc}
                            0 & 0 & 0 & 1 \\
                          \end{array}
                        \right)_n .
\end{eqnarray}
In this notation, the system of equations (\ref{meq}) takes the form
\begin{equation}\label{fC}
    \langle f^j(n),C\rangle \equiv \sum_{s \in S_{13}} V^j{}_k(n,s) c^k(n+s) = 0,
\end{equation}
where $j=1,2,3,4$, $n \in {\mathcal L}$, and
\begin{eqnarray}
  &&f^j(n)=\sum_{s \in S_{13}} V^j{}_k(n,s) \theta^k(n+s),\nonumber\\
  &&\langle f^j(n),e_k(n+s)\rangle=V^j{}_k(n,s).\label{fjn}
\end{eqnarray}
These relations can be rearranged to the basic system of equations
\begin{equation}\label{PnC}
    P(n)C = 0, \quad n \in {\mathcal L},
\end{equation}
where
\begin{equation}\label{Pnfaf}
    P(n) = [f^{\alpha}(n)]^\dag \otimes a^{\alpha}{}_\beta (n)f^\beta (n)
\end{equation}
is the Hermitian projection operator with the trace $tr[P(n)]=4$ and following properties:
\begin{equation}\label{Pn2}
   [P(n)]^2 = [P(n)]^\dag= P(n),
\end{equation}
\begin{equation}\label{anL}
    a(n) = [L(n)]^{-1}, \quad L^\alpha{}_\beta (n) =\left\langle f^\alpha,\left[f^{\beta}(n)\right]^\dag\right\rangle,
\end{equation}
where $\alpha , \beta=1,2,3,4$. The Hermitian $4\times 4$ matrices $L(n)$ and $a(n)$ at $n=(n_1,n_2,n_3,n_4)$ are defined by the following D-sets:
\begin{eqnarray}
   &&D_s[L(n)]=\left\{1+I_A+w_1^2+w_2^2+w_3^2+w_4^2,0,0,0,\right.\nonumber\\
   &&\left. -2w_4,0,0,0,0,2w_3w_4,2w_1w_4,2w_2w_4,0,0,0,0 \right\},\label{dsLn}
\end{eqnarray}
\begin{eqnarray}
   &&D_s[a(n)]=\frac{1}{\left|L(n)\right|}\left\{1+I_A+w_1^2+w_2^2+w_3^2+w_4^2,0,0,0,\right.\nonumber\\
   &&\left. 2w_4,0,0,0,0,-2w_3w_4,-2w_1w_4,-2w_2w_4,0,0,0,0\right\},\label{dsan}
\end{eqnarray}
where
\begin{eqnarray}
    I_A =&& 2\sum_{j=1}^6 \left|\bm{A}_j \right|^2 = 2\left(a_{12}^2+b_{12}^2+a_{13}^2+b_{13}^2+a_{21}^2+b_{21}^2\right. \nonumber\\ &&+a_{23}^2+b_{23}^2+a_{31}^2+b_{31}^2+a_{32}^2+b_{32}^2\nonumber \\
    &&+a_{42}^2+b_{42}^2+a_{43}^2+b_{43}^2+a_{51}^2+b_{51}^2\nonumber \\
    &&\left. +a_{53}^2 +b_{53}^2+a_{61}^2+b_{61}^2+a_{62}^2+b_{22}^2\right),\label{IA}
\end{eqnarray}
\begin{eqnarray}
  \left|L(n)\right|=&&I_A^2 + 2I_A\left(1+w_1^2+w_2^2+w_3^2+w_4^2\right)\nonumber\\
  &&+\left(1+w_1^2+w_2^2+w_3^2-w_4^2\right)^2.\label{dLn}
\end{eqnarray}
It is significant that, for a nonvanishing electromagnetic field ($I_A\neq 0$), the determinant $\left|L(n)\right|>0$ and hence equations (\ref{PnC})--(\ref{dLn}) are valid for any $n \in {\mathcal L}$.

\section{Projection operator of a system of homogeneous linear equations\label{project}}
Let $\mathcal{V}$ and $\mathcal{V}^\ast$ be a linear space (finite or infinite dimensional) and its dual. At given $\omega \in \mathcal{V}^\ast$, the linear homogeneous equation in $\bm{x} \in \mathcal{V}$
\begin{equation}\label{omx}
    \langle\omega,\bm{x}\rangle = 0
\end{equation}
can be transformed to the equivalent equation
\begin{equation}\label{alx}
    \alpha\bm{x} = 0
\end{equation}
where
\begin{equation}\label{aldyad}
    \alpha = \frac{\omega^\dag\otimes\omega}{\langle\omega,\omega^\dag\rangle}
\end{equation}
is the Hermitian projection operator (dyad) with the trace $tr\,\alpha=1$, and $\omega^\dag\in\mathcal{V}$. Let $U$ be the unit operator, i.e., $U\bm{x}=\bm{x}$ for any $\bm{x} \in \mathcal{V}$ and $\omega U = \omega$ for any $\omega \in \mathcal{V}^\ast$. The Hermitian projection operator $S=U-\alpha$ is the fundamental solution of (\ref{alx}), i.e., for any given $\bm{x}_0 \in \mathcal{V}$, $\bm{x}=S\bm{x}_0$ is a partial solution of (\ref{omx}) and (\ref{alx}).

Let now $\alpha$ and $\beta$ be Hermitian projection operators ($\alpha^\dag=\alpha^2=\alpha, \beta^\dag=\beta^2=\beta$) in $\mathcal{V}$. Providing the series
\begin{equation}\label{Aab}
    A=\alpha+\beta+\sum_{k=1}^{+\infty} \left[(\alpha\beta)^k\alpha-(\alpha\beta)^k+(\beta\alpha)^k\beta-(\beta\alpha)^k\right]
\end{equation}
is convergent, it defines the Hermitian projection operator with the following properties
\begin{eqnarray}
  A^\dag=A^2=A,\quad \alpha A = A \alpha=\alpha,\nonumber\\
  \beta A = A \beta=\beta,\quad tr\,A=tr\,\alpha+ tr\,\beta. \label{Aprop}
\end{eqnarray}
Hence, the system of equations in $\bm{x}\in\mathcal{V}$
\begin{equation}\label{axbx}
    \alpha\bm{x}=0, \quad \beta\bm{x}=0
\end{equation}
reduces to one equation $A\bm{x}=0$ and has the fundamental solution $S=U-A$. The operator $A$ will be designated the projection operator of the system (\ref{axbx}). The trace $tr\,\alpha$ of the projection operator $\alpha$ specifies the dimension of the image $\alpha(\mathcal{V})$ of $\mathcal{V}$ under the mapping $\alpha$. It is significant that the relations (\ref{Aab}) and (\ref{Aprop}) are valid for any values of integers $tr\,\alpha$ and $tr\,\beta$. This enables us to extend this approach to systems with any (finite or infinite) number of homogeneous linear equations. To this end, we transform~(\ref{Aab}) to the following expression~\cite{bian04}
\begin{equation}
    A=(\alpha-\alpha\beta\alpha)^{-}(U-\beta)+(\beta-\beta\alpha\beta)^{-}(U-\alpha),\label{Apseu}
\end{equation}
where $(\alpha-\alpha\beta\alpha)^{-}$ is the pseudoinverse operator with the following properties
\begin{eqnarray}
   &&(\alpha-\alpha\beta\alpha)^{-}(\alpha-\alpha\beta\alpha)=(\alpha-\alpha\beta\alpha)(\alpha-\alpha\beta\alpha)^{-}=\alpha,\nonumber\\
   &&\alpha(\alpha-\alpha\beta\alpha)^{-}=(\alpha-\alpha\beta\alpha)^{-}\alpha=(\alpha-\alpha\beta\alpha)^{-},\nonumber\\
   &&\sum_{k=1}^{+\infty}(\alpha\beta)^k=(\alpha-\alpha\beta\alpha)^{-}\beta.\label{pseu3}
\end{eqnarray}
The similar relations for $(\beta-\beta\alpha\beta)^{-}$ can be obtained from (\ref{pseu3}) by the replacement $\alpha\leftrightarrow\beta$. Numerical implementation of the pseudoinversion reduces to the inversion of $(tr\,\alpha)\times (tr\,\alpha)$ matrix for $(\alpha-\alpha\beta\alpha)^{-}$ and $(tr\,\beta)\times (tr\,\beta)$ matrix for $(\beta-\beta\alpha\beta)^{-}$.

In~\cite{bian04}, we have proposed a technique based on the use of (\ref{Apseu}) to find the fundamental solution of  the system~(\ref{PnC}). In the current series of papers, we present the advanced version of this technique based on a fractal expansion of the system of equations taking into account (the following paper) and on the use of $A$ (\ref{Aab}) expressed as
\begin{equation}\label{Aad}
    A=\alpha+\delta, \quad \delta=(\beta-\alpha)\gamma(\beta-\alpha),
\end{equation}
where
\begin{equation}\label{gam}
    \gamma=\beta+\sum_{k=1}^{+\infty}(\beta\alpha\beta)^k=(\beta-\beta\alpha\beta)^{-},
\end{equation}
$\alpha,\beta,\delta$, and $A$ are projection operators, $\alpha,\beta,\gamma,\delta$, and $A$ are Hermitian operators interrelated as
\begin{eqnarray}
  &&\beta\gamma=\gamma\beta=\gamma,\quad \beta\alpha\gamma=\gamma\alpha\beta=\gamma-\beta,\nonumber\\
  &&\alpha\delta=\delta\alpha=0,\quad \beta\delta=\beta-\beta\alpha,\quad \delta\beta=\beta-\alpha\beta,\nonumber\\
  &&\alpha A=A\alpha=\alpha, \quad \beta A=A\beta=\beta, \quad \delta A=A\delta=\delta. \label{abgd}
\end{eqnarray}
In the frame of this approach, calculation of all pseudoinverse operators in use reduces to the inversion of $4\times 4$ matrices.

\section{Recurrent algorithm\label{recur}}
In this section, we present the recurrent algorithm to find the projection operator of the basic system of equations~(\ref{PnC}).
\subsection{Product $P(m)P(n)$}
It follows from Eq.~(\ref{Pnfaf}) that
\begin{equation}\label{PmPn}
    P(m)P(n)=\left[f^i(m)\right]^{\dag}\otimes\left[a(m)N(m,n)a(n)\right]^i{}_j f^j(n),
\end{equation}
where
\begin{equation}\label{Nij}
    N^i{}_j(m,n)=\left\langle f^i(m),\left[f^j(n)\right]^{\dag}\right\rangle,\quad i,j=1,2,3,4,
\end{equation}
$N(n,n)\equiv L(n)$ (\ref{anL}). At any given n, Eq.~(\ref{meq}) relates the Fourier amplitude $c(n)$ only with 12 amplitudes $c(n+s)$, where $g_{4d}(s)=1$. In consequence of this, $N(m,n)\equiv 0$ at $g_{4d}(n-m)>2$. Substitution of (\ref{fjn}) in (\ref{Nij}) at $n=m+s$ gives
\begin{eqnarray}
  N^{\dag}(n,m)=N(m,n)&=&L(m) \text{ for } n=m,\nonumber\\
                      &=&N_1(m,s) \text{ for } g_{4d}(s)=1,\nonumber\\
                      &=&N_2(s)\Gamma_0 \text{ for } g_{4d}(s)=2.
\end{eqnarray}
The D-sets of 12 matrices $N_1(m,s)$ and the table of 56 scaler coefficients $N_2(s)$ will be presented in the third paper of this series.

\subsection{Sublattices}
The Hermitian operator $\mathcal P$ of the system of equations~(\ref{PnC}), by definition, has the following properties
\begin{equation}\label{Pdag}
    \mathcal{P}^\dag=\mathcal{P}^2=\mathcal{P}, \quad P(n)\mathcal{P}=\mathcal{P}P(n)=P(n)
\end{equation}
for any $n\in{\mathcal L}$. Let us seek it in the form of a series
\begin{equation}\label{Pro}
    \mathcal{P}=\sum_{k=0}^{+\infty}\sum_{n\in\mathcal{F}_k}\rho_k(n),
\end{equation}
where $\mathcal{F}_k$ are the lattices satisfying the conditions
\begin{equation}\label{FkL}
    \bigcup_{k=0}^{+\infty}\mathcal{F}_k=\mathcal{L},\quad \mathcal{F}_j\bigcap\mathcal{F}_k=\emptyset,\quad j\neq k,
\end{equation}
and $\rho_k(n)$ are Hermitian projection operators satisfying the relations
\begin{equation}\label{rok}
    \rho_k^{\dag}(n)=\rho_k^2(n)=\rho_k(n), \quad tr[\rho_k(n)]=4, \quad n \in \mathcal{L},
\end{equation}
\begin{equation}\label{romn}
    \rho_k(m)\rho_l(n)=0 \text{ if } k\neq l \text{ or (and) } m\neq n,
\end{equation}
\begin{equation}\label{ro0}
    \rho_0(n)=P(n), \quad n\in \mathcal{F}_0.
\end{equation}
There exist various ways to split the lattice $\mathcal L$ into sublattices $\mathcal{F}_k$ to fulfil conditions (\ref{FkL}) and (\ref{romn}), one of them will be described in the second paper of this series. Providing these conditions are met, substitution of
\begin{equation}\label{albero}
    \alpha=\sum_{j=0}^{k-1}\sum_{n\in \mathcal{F}_j}\rho_j(n)\equiv\mathcal{P}_{k-1}, \quad \beta=P(m), \quad m\in \mathcal{P}_{k}
\end{equation}
into Eqs.~(\ref{Aad}) and (\ref{gam}) results in $\rho_k(m)=\delta$ (\ref{Aad}).

All operators $\rho_k(m)$ have the same trace $tr\left[\rho_k(m)\right]=4$ and can be written as
\begin{equation}\label{rokm}
    \rho_k(m)=\left[F_k^{\alpha}(m)\right]^{\dag}\otimes \left[A_k(m)\right]^{\alpha}{}_{\beta}F_k^{\beta}(m), \quad m\in \mathcal{F}_k,
\end{equation}
where $k=0,1,...$, and
\begin{eqnarray}
  F_0^{\alpha}(m)\equiv f^{\alpha}(m), \quad A_0(m)\equiv a(m),\nonumber\\
  F_k^{\alpha}(m)=f^{\alpha}(m)-g_k^{\alpha}(m), \quad k=1,2,...;\label{Fkm}
\end{eqnarray}
\begin{eqnarray}
  g_k^{\alpha}(m)&&\equiv f^{\alpha}(m)\mathcal{P}_{k-1}=\sum_{j=0}^{k-1}\sum_{n\in \mathcal{F}_j} \left[C_{kj}(m,n)\right]^{\alpha}{}_{\beta}f^{\beta}(n),\nonumber\\
  \left[g_k^{\alpha}(m)\right]^{\dag}&&\equiv\mathcal{P}_{k-1}\left[f^{\alpha}(m)\right]^{\dag}\nonumber\\
  &&=\sum_{j=0}^{k-1}\sum_{n\in \mathcal{F}_j} \left[f^{\beta}(n)\right]^{\dag} \left[C_{jk}(n,m)\right]^{\beta}{}_{\alpha}; \label{gkm}
\end{eqnarray}
\begin{eqnarray}
  A_k(m)=\left[L(m)-G_k(m)\right]^{-1},\nonumber\\
  G_k(m)=\sum_{j=0}^{k-1}\sum_{n\in \mathcal{F}_j} C_{kj}(m,n)N(n,m).\label{AkGk}
\end{eqnarray}

\subsection{Recurrent relations}
The $4\times 4$ matrices $C_{kj}(m,n)$ and $C_{jk}(n,m)$ are related as
\begin{equation}\label{CjkCkj}
    C_{jk}(n,m)=\left[C_{kj}(m,n)\right]^{\dag},
\end{equation}
where $m\in \mathcal{F}_k$, $k=1,2,...$; $n\in \mathcal{F}_j$, $j=0,1,...,k-1$. The relationship between one-forms $f^{\alpha}(m)$ and $F_j^{\beta}(n)$ is described by $4\times 4$ matrix $D_{kj}(m,n)$ as
\begin{equation}\label{fF}
    f^{\alpha}(m)\rho_j(n)=\left[D_{kj}(m,n)\right]^{\alpha}{}_{\beta}F_j^{\beta}(n),
\end{equation}
where
\begin{eqnarray}
  D_{kj}(m,n)=&&N(m,n)A_j(n)\nonumber\\
  &&-\sum_{i=0}^{j-1}\sum_{p\in \mathcal{F}_i} N(m,p)C_{ij}(p,n)A_j(n),\label{Dkj}
\end{eqnarray}
where $m\in \mathcal{F}_k, k=2,3,...;n\in \mathcal{F}_j, j=1,...,k-1.$

The families of matrices $C_{kj}(m,n)$ and $D_{kj}(m,n)$ are defined by Eqs. (\ref{CjkCkj}), (\ref{Dkj}), and the recurrent relations:
\begin{equation}\label{C10}
    C_{10}(m,n)=N(m,n)a(n), \quad m\in \mathcal{F}_1, \quad n\in \mathcal{F}_0;
\end{equation}
\begin{eqnarray}
  &&C_{k0}(m,n)=N(m,n)a(n)-\sum_{j=1}^{k-1}\sum_{p\in \mathcal{F}_j}D_{kj}(m,p)C_{j0}(p,n),\nonumber\\
  &&m\in \mathcal{F}_k,\quad k=2,3,...,\quad n\in \mathcal{F}_0;\label{Ck0}
\end{eqnarray}
\begin{eqnarray}
  &&C_{ki}(m,n)=D_{ki}(m,n)-\sum_{j=i+1}^{k-1}\sum_{p\in \mathcal{F}_j}D_{kj}(m,p)C_{ji}(p,n),\nonumber\\
  &&m\in \mathcal{F}_k,\quad k=3,4,...,\quad n\in \mathcal{F}_i \quad i=1,...,k-2;\label{Cki}
\end{eqnarray}
\begin{equation}\label{Ckk1}
    C_{k\,k-1}(m,n)=D_{k\,k-1}(m,n), \quad k=2,3,... .
\end{equation}
From Eqs.~(\ref{Fkm})--(\ref{Ckk1}) it follows
\begin{equation}\label{Akdag}
    \left[A_k(m)\right]^{\dag}=A_k(m), \quad \left[G_k(m)\right]^{\dag}=G_k(m),
\end{equation}
that is, $G_k(m)$ and $A_k(m)$ are Hermitian matrices with real D-sets.

In numerical calculations, the projection operator $\rho_k(m)$ (\ref{rokm}) is represented by its components
\begin{equation}\label{{Rkrok}}
    \left[R_k(m',m,n')\right]^{\mu}{}_{\nu}=\left\langle\theta^{\mu}(m'),\rho_k(m)e_{\nu}(n')\right\rangle,
\end{equation}
where $m', n' \in {\mathcal L}; \mu, \nu=1,2,3,4.$ They can be conveniently treated as elements of $4\times 4$ matrices
\begin{equation}\label{RkA}
    R_k(m',m,n')=\left[\Phi_k(m,m')\right]^{\dag}A_k(m)\Phi_k(m,n'),
\end{equation}
where $\left[\Phi_k(m,n')\right]^{\alpha}{}_{\mu}\equiv\left\langle F_k^{\alpha}(m),e_{\mu}(n')\right\rangle,$ and
\begin{eqnarray}
  \Phi_k(m,n')=&&V(m,n'-m)\nonumber\\
  &&-\sum_{j=0}^{k-1}\sum_{n\in \mathcal{F}_j} C_{kj}(m,n)V(n,n'-n).\label{Phik}
\end{eqnarray}
The Hermitian matrix $A_k(m)$ and the set of matrices $\Phi_k(m,n')$ uniquely define the projection operator $\rho_k(m)$.

\subsection{Fundamental solution}
The fundamental solution of Eq.~(\ref{PnC}), i.e., the operator of projection onto the solution subspace of the multispinor space $V_C$, has the form
\begin{equation}\label{SUP}
    \mathcal{S}=\mathcal{U}-\mathcal{P},
\end{equation}
where $\mathcal{U}$ is the unit operator in $V_C$, which can be written as
\begin{equation}\label{UVC}
    \mathcal{U}=\sum_{n\in \mathcal{L}} I(n),\quad I(n)=e_j(n)\otimes\theta^j(n), \quad tr[I(n)]=4.
\end{equation}
For any $C_0\in V_C$, $C=\mathcal{S}C_0$ is a partial solution of Eq.~(\ref{PnC}), i.e., the function $\Psi$~(\ref{sol1}) with the set of Fourier amplitudes $\{c(n), n \in \mathcal{L}\}=\mathcal{S}C_0$  satisfies the Dirac equation~(\ref{Dirac}) for the problem under consideration.

\section{Conclusion}
A system of homogeneous linear equations in a finite--dimensional space or an infinite--dimensional space is characterized by the Hermitian projection operator of the system, directly coupled with the fundamental solution --- the operator of projection onto the subspace of solutions. The relations presented in section~\ref{project} provide convenient means to find these operators by making use a recurrent algorithm and the pseudoinversion.  In the frame of this general approach, the fundamental solution of the Dirac equation for an electron in the electromagnetic field with four--dimensional periodicity is obtained.

\appendix*
\section{}
\subsection{Dirac basis for the linear space of $4\times 4$ matrices}
Let us enumerate 16 Dirac matrices, forming a basis for the linear space of $4\times 4$ matrices, by taking into account both interrelations between $2\times 2$ blocks of each matrix and interrelations between elements of each nonzero $2\times 2$ block as follows
\begin{equation*}
    \Gamma_0=\Gamma_{0000}=\left(
                            \begin{array}{cccc}
                              1 & 0 & 0 & 0 \\
                              0 & 1 & 0 & 0 \\
                              0 & 0 & 1 & 0 \\
                              0 & 0 & 0 & 1 \\
                            \end{array}
                          \right)=U,
\end{equation*}
\begin{equation*}
    \Gamma_1=\Gamma_{0001}=\left(
                            \begin{array}{cccc}
                              1 & 0 & 0 & 0 \\
                              0 & -1 & 0 & 0 \\
                              0 & 0 & 1 & 0 \\
                              0 & 0 & 0 & -1 \\
                            \end{array}
                          \right)=\Sigma_3,
\end{equation*}
\begin{equation*}
    \Gamma_2=\Gamma_{0010}=\left(
                            \begin{array}{cccc}
                              0 & 1 & 0 & 0 \\
                              1 & 0 & 0 & 0 \\
                              0 & 0 & 0 & 1 \\
                              0 & 0 & 1 & 0 \\
                            \end{array}
                          \right)=\Sigma_1,
\end{equation*}
\begin{equation*}
    \Gamma_3=\Gamma_{0011}=\left(
                            \begin{array}{cccc}
                              0 & -i & 0 & 0 \\
                              i & 0 & 0 & 0 \\
                              0 & 0 & 0 & -i \\
                              0 & 0 & i & 0 \\
                            \end{array}
                          \right)=\Sigma_2,
\end{equation*}
\begin{equation*}
    \Gamma_4=\Gamma_{0100}=\left(
                            \begin{array}{cccc}
                              1 & 0 & 0 & 0 \\
                              0 & 1 & 0 & 0 \\
                              0 & 0 & -1 & 0 \\
                              0 & 0 & 0 & -1 \\
                            \end{array}
                          \right)=\gamma_4=\alpha_4,
\end{equation*}
\begin{equation*}
    \Gamma_5=\Gamma_{0101}=\left(
                            \begin{array}{cccc}
                              1 & 0 & 0 & 0 \\
                              0 & -1 & 0 & 0 \\
                              0 & 0 & -1 & 0 \\
                              0 & 0 & 0 & 1 \\
                            \end{array}
                          \right)=\tau_3,
\end{equation*}
\begin{equation*}
    \Gamma_6=\Gamma_{0110}=\left(
                            \begin{array}{cccc}
                              0 & 1 & 0 & 0 \\
                              1 & 0 & 0 & 0 \\
                              0 & 0 & 0 & -1 \\
                              0 & 0 & -1 & 0 \\
                            \end{array}
                          \right)=\tau_1,
\end{equation*}
\begin{equation*}
    \Gamma_7=\Gamma_{0111}=\left(
                            \begin{array}{cccc}
                              0 & -i & 0 & 0 \\
                              i & 0 & 0 & 0 \\
                              0 & 0 & 0 & i \\
                              0 & 0 & -i & 0 \\
                            \end{array}
                          \right)=\tau_2,
\end{equation*}
\begin{equation*}
    \Gamma_8=\Gamma_{1000}=\left(
                            \begin{array}{cccc}
                              0 & 0 & -1 & 0 \\
                              0 & 0 & 0 & -1 \\
                              -1 & 0 & 0 & 0 \\
                              0 & -1 & 0 & 0 \\
                            \end{array}
                          \right)=\gamma_5,
\end{equation*}
\begin{equation*}
    \Gamma_9=\Gamma_{1001}=\left(
                            \begin{array}{cccc}
                              0 & 0 & 1 & 0 \\
                              0 & 0 & 0 & -1 \\
                              1 & 0 & 0 & 0 \\
                              0 & -1 & 0 & 0 \\
                            \end{array}
                          \right)=\alpha_3
\end{equation*}
\begin{equation*}
    \Gamma_{10}=\Gamma_{1010}=\left(
                            \begin{array}{cccc}
                              0 & 0 & 0 & 1 \\
                              0 & 0 & 1 & 0 \\
                              0 & 1 & 0 & 0 \\
                              1 & 0 & 0 & 0 \\
                            \end{array}
                          \right)=\alpha_1,
\end{equation*}
\begin{equation*}
    \Gamma_{11}=\Gamma_{1011}=\left(
                            \begin{array}{cccc}
                              0 & 0 & 0 & -i \\
                              0 & 0 & i & 0 \\
                              0 & -i & 0 & 0 \\
                              i & 0 & 0 & 0 \\
                            \end{array}
                          \right)=\alpha_2,
\end{equation*}
\begin{equation*}
    \Gamma_{12}=\Gamma_{1100}=\left(
                            \begin{array}{cccc}
                              0 & 0 & i & 0 \\
                              0 & 0 & 0 & i \\
                              -i & 0 & 0 & 0 \\
                              0 & -i & 0 & 0 \\
                            \end{array}
                          \right)=\tau_4,
\end{equation*}
\begin{equation*}
    \Gamma_{13}=\Gamma_{1101}=\left(
                            \begin{array}{cccc}
                              0 & 0 & -i & 0 \\
                              0 & 0 & 0 & i \\
                              i & 0 & 0 & 0 \\
                              0 & -i & 0 & 0 \\
                            \end{array}
                          \right)=\gamma_3,
\end{equation*}
\begin{equation*}
    \Gamma_{14}=\Gamma_{1110}=\left(
                            \begin{array}{cccc}
                              0 & 0 & 0 & -i \\
                              0 & 0 & -i & 0 \\
                              0 & i & 0 & 0 \\
                              i & 0 & 0 & 0 \\
                            \end{array}
                          \right)=\gamma_1,
\end{equation*}
\begin{equation*}
    \Gamma_{15}=\Gamma_{1111}=\left(
                            \begin{array}{cccc}
                              0 & 0 & 0 & -1 \\
                              0 & 0 & 1 & 0 \\
                              0 & 1 & 0 & 0 \\
                              -1 & 0 & 0 & 0 \\
                            \end{array}
                          \right)=\gamma_2.
\end{equation*}
At the presented numeration order, the structural information on each matrix $\Gamma_{\nu}=\Gamma_{MNmn}$ is enclosed in its number which is written above in decimal notation ($\nu$) and binary notation ($MNmn$) with four binary digits for any $\nu=0,...,15$, i.e.,
\begin{equation}\label{nu842}
    \nu=8 M + 4 N +2 m +n.
\end{equation}
Commonly used notation to the right of each matrix is given for convenience.

To reconstruct the matrix from its number, first we calculate the nonzero element
\begin{equation}\label{bnu}
    b_{\nu}=b_{MNmn}=i^{M N + m n}(-1)^{(1-M)m n + M(1+N+m+n)}
\end{equation}
of the first matrix row, which is situated in $2M+m+1$ column. The binary digits $m$ and $n$ define the structure of $2\times 2$ block $X$ containing this element as follows
\begin{equation*}
    X=\left(
                 \begin{array}{cc}
                   b_{MN0n} & 0 \\
                   0 & (-1)^n b_{MN0n}\\
                 \end{array}
               \right) \text{ for } m=0,
\end{equation*}
\begin{equation*}
    X=\left(
                 \begin{array}{cc}
                   0 & b_{MN1n} \\
                   (-1)^n b_{MN1n} & 0\\
                 \end{array}
               \right) \text{ for } m=1.
\end{equation*}
Finally, the digits $M$ and $N$ uniquely define
\begin{equation*}
    \Gamma_{\nu}=\Gamma_{MNmn}=\left(
                                \begin{array}{cc}
                                  A & B \\
                                  C & D \\
                                \end{array}
                              \right)
\end{equation*}
in terms of its $2\times 2$ blocks $A,B,C$ and $D$ as
\begin{equation*}
    B=C=\left(
                 \begin{array}{cc}
                   0 & 0 \\
                   0 & 0\\
                 \end{array}
               \right), D=(-1)^N A, A=X \text{ for } M=0,
\end{equation*}
\begin{equation*}
    A=D=\left(
                 \begin{array}{cc}
                   0 & 0 \\
                   0 & 0\\
                 \end{array}
               \right), C=(-1)^N B, B=X \text{ for } M=1.
\end{equation*}

The matrix product $\Gamma_{\lambda}\Gamma_{\mu}$ at all values of $\lambda$ and $\mu$ can be written as
\begin{equation*}
    \Gamma_{\lambda}\Gamma_{\mu}=f_{\lambda\mu}\Gamma_{\nu}.
\end{equation*}
It can be described by $16\times 16$ multiplication tables of $\nu$ and $f_{\lambda\mu}$ values depending on $\lambda, \mu=0,...,15$. The presented numeration provides a simple way to describe this multiplication rule without recourse to tables. By using (\ref{nu842}), (\ref{bnu}) and the binary  forms $GHgh$ and $JKjk$ of numbers $\lambda$ and $\mu$, i.e.,
\begin{equation*}
    \lambda=8G+4H+2g+h, \quad \mu=8J+4K+2j+k,
\end{equation*}
we obtain
\begin{equation*}
 M=|G-J|, N=|H-K|, m=|g-j|, n=|h-k|,
\end{equation*}
\begin{eqnarray*}
  f_{\lambda\mu}&=&f[(G,H,g,h),(J,K,j,k)] \\
  &=&\frac{b_{GHgh}b_{JKjk}}{b_{MNmn}}(-1)^{GK+gk} = i^{GK+JH+gk+jh}(-1)^Z,
\end{eqnarray*}
where
\begin{eqnarray*}
 Z=&&G K(1-J-H)+J H(G+K)\\
   &&+(G j+J g)(1-h-k)+G k(1-g)\\
   &&+J h(1-j)+g k(1-j-h)+j h(g+k).
\end{eqnarray*}

\subsection{Dirac set of $4\times 4$ matrix}
In terms of Eq.~(\ref{nu842}) any $4\times 4$ matrix $A$ can be written
\begin{equation*}
    A=\sum_{\nu=0}^{15}A_{\nu}\Gamma_{\nu}\equiv\sum_{M,N,m,n=0}^1 A_{MNmn}\Gamma_{MNmn},
\end{equation*}
where $A_{\nu}\equiv A_{MNmn}=\frac14 tr(A\Gamma_{\nu})$, and $tr\, A=4 A_0$. To single out the specific basis used in this expansion, the set of coefficients with decimal $\{A_{\nu}\}$ or binary $\{A_{MNmn}\}$ indices is called in this article the Dirac set of matrix $A$, briefly, D-set of $A$, and it is denoted $D_s(A)$. This approach is of particular assistance in solving the system of Eqs.~(\ref{meq}). It is best suited to the structure of its matrix coefficients, accelerates numerical calculations and reduces data files.

It should be emphasized that all major matrix operations can be performed directly with D-sets, i.e., without matrix form retrieval. In particular, the function $P_D$, describing the matrix product $C=A B$ in terms of D-sets, is given by
\begin{equation*}
    D_s(C)=P_D[D_s(A),D_s(B)],
\end{equation*}
\begin{eqnarray*}
  C_{MNmn}=&&\sum_{G,H,g,h=0}^1 A_{GHgh}B_{JKjk}\\
           &&\times f[(G,H,g,h),(J,K,j,k)],
\end{eqnarray*}
where
\begin{equation*}
  J=|M-G|, K=|N-H|, j=|m-g|, k=|n-h|.
\end{equation*}
The map $A\mapsto D_s(A)$ and its inverse $D_s(A)\mapsto A$ are linear, and $D_s(A^{\dag})=[D_s(A)]^\ast$, i.e., D-set of a Hermitian matrix is real.

Let us assume that D-set of matrix $A$ has the form
\begin{equation*}
    D_s(A)=\{a,b,c,d,e,f,g,h,s,t,u,v,w,x,y,z\}.
\end{equation*}
Then the coefficients $I_1, I_2, I_3$ and $I_4$ of its characteristic equation
\begin{equation*}
    \lambda^4-I_1\lambda^3+I_2\lambda^2-I_3\lambda+I_4=0
\end{equation*}
are given by the expressions:
\begin{eqnarray*}
  I_4 = \left[(a - e)^2 - (b - f)^2 - (c - g)^2 - (d - h)^2\right] \\
  \times\left[(a + e)^2 - (b + f)^2 - (c + g)^2 - (d + h)^2\right]\\
  + 4(-s w + t x + u y + v z)^2\\
  + \left(s^2 - t^2 - u^2 - v^2 - w^2 + x^2 + y^2 + z^2\right)^2\\
  -2\left[\left(b^2 - f^2\right)\left(s^2 + t^2 - u^2 - v^2 + w^2 + x^2 - y^2 - z^2\right)\right. \\
  + \left(c^2 - g^2\right)\left(s^2 - t^2 + u^2 - v^2 + w^2 - x^2 + y^2 - z^2\right) \\
  + \left(d^2 - h^2\right)\left(s^2 - t^2 - u^2 + v^2 + w^2 - x^2 - y^2 + z^2\right) \\
  \left. + \left(a^2 - e^2\right)\left(s^2 + t^2 + u^2 + v^2 + w^2 + x^2 + y^2 + z^2\right)\right] \\
  - 8\left[(d g - c h)(s x - t w) + (a b - e f)(s t + w x)\right.\\
   + (d f - b h)(u w - s y) + (d e - a h)(u x - t y) \\
  + (a c - e g)(s u + w y) + (b c - f g)(t u + x y) \\
  + (b g - c f )(v w - s z) + (a g - c e)(v x - t z)\\
   + (b e - a f)(v y - u z) + (a d - e h)(s v + w z)\\
   \left. + (b d - f h)(t v + x z) + (c d - g h)(u v + y z)\right],
\end{eqnarray*}
\begin{eqnarray*}
  I_3=&&4a\left(a^2-I_0\right)\\
      &&+8\left[c e g + d e h - c s u - d s v + h u x - g v x \right.\\
      &&+b(e f - s t - w x)+y(f v - h t - c w)\\
      &&\left. +z(g t - d w - f u)\right],
\end{eqnarray*}
\begin{equation*}
    I_2=6a^2-2I_0,
\end{equation*}
\begin{equation*}
   I_1=4a,
\end{equation*}
where
\begin{eqnarray*}
  I_0=&&b^2+c^2+d^2+e^2+f^2+g^2+h^2+s^2\\
  &&+t^2+u^2+v^2+w^2+x^2+y^2+z^2.
\end{eqnarray*}
Here, $I_1=tr\, A$ and $I_4=|A|$ are the trace and the determinant of $A$, $I_3=tr\,\overline{A}$ is the trace of the adjoint matrix $\overline{A}$ defined by the equation $A\overline{A}=\overline{A}A=|A|\Gamma_0$.The Hamilton--Cayley theorem provides the relation
\begin{equation*}
    \overline{A}=I_3\Gamma_0 - I_2 A + I_1 A^2 - A^3.
\end{equation*}
Hence, the D-sets of the adjoint matrix $\overline{A}$ and the inverse matrix $A^{-1}$ (assuming $I_4\neq 0$) are defined by the relations
\begin{eqnarray*}
  D_s(\overline{A})=&&I_3D_s(\Gamma_0) - I_2D_s(A)\\
   &&+P_D[D_s(A^2),I_1D_s(\Gamma_0)-D_s(A)],\\
  D_s(A^{-1})=&&D_s(\overline{A})/I_4,
\end{eqnarray*}
where $D_s(\Gamma_0)=\{1,0,0,0,0,0,0,0,0,0,0,0,0,0,0,0\}$, and
\begin{eqnarray*}
  D_s(A^2)=P_D[D_s(A),D_s(A)]=(a^2+I_0)D_s[\Gamma_0]  \\
  +2\left\{0,a b + e f - s t - w x, \right.\\
   a c + e g - s u - w y, a d + e h - s v - w z,\\
   a e + b f + c g + d h, b e + a f +v y - u z,\\
   c e + a g - v x + t z, d e + a h + u x - t y,\\
   a s - b t - c u - d v, a t - b s - h y + g z,\\
   a u - c s + h x -f z, a v - s d - g x + f y,\\
   a w - b x - c y - d z, h u - g v - b w + a x,\\
   \left. f v - h t - c w + a y, g t - f u - d w + a z \right\}.
\end{eqnarray*}

\bibliography{Borzdov1}
\end{document}